\documentclass[useAMS,usenatbib,usegraphicx]{mn2e}

\title[The VLSS Redux]{The Very Large Array Low-frequency Sky Survey Redux (VLSSr)}
\author[W. M. Lane et al.]
  { W.~M.~Lane,$^{1}$\thanks{E-mail: wendy.peters@nrl.navy.mil}
    W.~ D.~Cotton,$^{2}$  S.~van~Velzen,$^{3}$ T.~E.~Clarke,$^{1}$ \\
    \newauthor
     N.~E.~Kassim,$^{1}$, J.~F.~Helmboldt,$^{1}$ T.~J.~W.~Lazio,$^{4}$ and A.~S. Cohen,$^{5}$ \\
    $^1$Naval Research Laboratory, Code 7213, 4555 Overlook Ave SW, Washington, DC 20375 \\
    $^2$National Radio Astronomy Observatory, 520 Edgemont Drive, Charlottesville, VA, 22903 \\
    $^3$Department of Astrophysics/IMAPP, Radboud University, P.O. Box 9010, 6500 GL Nijmegen, The Netherlands \\
    $^4$Jet Propulsion Laboratory, California Institute of Technology, M/S 138-308, 4800 Oak Grove Dr, Pasadena, CA  91109 \\
    $^5$The Johns Hopkins University Applied Physics Laboratory, 11100 Johns Hopkins Rd, Laurel, MD 20723
}

\begin{document}

\date{accepted 2014 February 5}


\maketitle

\label{firstpage}

\begin{abstract}

We present the results of a recent re-reduction of the data from the
Very Large Array (VLA) Low-frequency Sky Survey (VLSS). We used the
VLSS catalog as a sky model to correct the ionospheric distortions in
the data and create a new set of sky maps and corresponding catalog at
73.8 MHz. The VLSS Redux (VLSSr) has a resolution of $75$ arcsec, and
an average map RMS noise level of $\sigma \sim 0.1$ Jy beam$^{-1}$.
The clean bias is $0.66 \times \sigma$ and the theoretical largest
angular size is $36$ arcmin.  Six previously unimaged fields are
included in the VLSSr, which has an unbroken sky coverage over 9.3 sr
above an irregular southern boundary. The final catalog includes
92,964 sources.  The VLSSr improves upon the original VLSS in a number
of areas including imaging of large sources, image sensitivity, and
clean bias; however the most critical improvement is the replacement
of an inaccurate primary beam correction which caused source flux
errors which vary as a function of radius to nearest pointing center
in the VLSS.

\end{abstract}


\section{Introduction}
Sky surveys provide an invaluable tool in many areas of astronomy.
The combination of data from surveys at a range of wavelengths allows
the study of the statistical properties of large source samples to
create a more complete understanding of the physical processes which
drive the sources.  For example, surveys can be used to mitigate
cosmic variance and allow the study of large samples of galaxies,
their properties, and their evolution over time
~\citep[eg.][]{Kochanek2012}. The large survey source samples can also
be used to find objects with rare or unusual properties
~\citep[eg.][]{Fan2002}.  Surveys at radio wavelengths are
particularly important to generate samples for study of active galactic nuclei (AGN) activity
and the star formation history of the Universe un-obscured by dust
~\citep[eg.][]{Archibald2001,Seymour2008}.

Existing radio surveys, including but not limited to the Australia
Telescope 20 GHz Survey ~\citep[AT20G:][]{AT20G}, the 5 GHz
Parkes-MIT-NRAO survey~\citep[PMN:][]{PMN}, the 1400 MHz NRAO VLA Sky
Survey ~\citep[NVSS:][]{NVSS}, the 843 MHz Sydney University Molonglo
Sky Survey ~\citep[SUMSS:][]{SUMSS}, the 330 MHz Westerbork Northern
Sky Survey ~\citep[WENSS:][]{WENSS}, the 178 MHz 6th Cambridge Survey
~\citep[6C:][]{6C}, the 150 MHz TIFR GMRT Sky Survey
~\citep[TGSS:][]{TGSS}, and the 38 MHz 8th Cambridge survey
~\citep[8C:][]{8C}, complement each other to offer a rich and complex
view of the sky across the radio wavelength regime.

Between 2001 and 2007, the Very Large Array (VLA) was used to make a
complete survey of the $3\pi$ sr of sky above declinations $\delta >
-30\degr$ at a frequency of 74 MHz.  The data were reduced to provide
a publicly available catalog and set of maps, which were released as
the VLA Low-frequency Sky Survey \citep[VLSS:][]{VLSS1}.  The VLSS has
a resolution of $80$ arcsec and an average RMS map sensitivity of
$\sigma \sim 0.130$ Jy beam$^{-1}$.  The final catalog and images are
complete over $95$ per cent of the observed sky area.

The reduction of the VLSS data was limited by the lack of an existing
sky model at a comparable frequency. At 74 MHz, ionospheric phase
fluctuations can cause apparent offsets in the positions of sources
which vary with time and across the field of view.  It is possible to
correct these by comparing the observed data to a reference grid of
known source positions and calculating position-dependent phase
corrections for the data.  For the VLSS, the reference grid was
created by extrapolating the flux densities of sources from the 1400
MHz NVSS to 74 MHz, using an assumed standard spectral index of
$\alpha = -0.7$\footnote{where $S_{\nu} \propto \nu^{\alpha}$}.  As a
result of errors introduced by this extrapolation, some fraction of
the data could not be corrected for ionospheric phase errors and were
lost completely, while other data were not optimally corrected,
introducing errors into the final maps.

Additionally, due to limitations in then-available automated flagging
procedures for radio-frequency interference (RFI), all short
baselines, where the RFI is typically strongest, were excluded from
the VLSS images. This halved the largest angular size imaged.  Every
eighth frequency channel was also removed due to internally generated
RFI~\citep{VLA74}.  Both of these steps improved the overall image
quality, but further reduced the amount of data included in the
images~\citep{VLSSr1}. Finally, the primary beam model used to correct
the VLSS was inaccurate, introducing position dependent flux errors.

Despite its limitations, the VLSS has served as an important
low-frequency anchor point for multi-wavelength studies of a broad
range of Galactic and extra-Galactic sources
~\citep[eg.][]{Brunetti2008, Kothes2008, Argo2013}.  It has provided a
low-frequency comparison point for other sky surveys
~\citep[eg.][]{Bernardi2013}, and a global sky model that can be used
for simulations ~\citep[eg.][]{Moore2013} and preliminary calibration
of other low-frequency instruments such as the Low-Frequency Array
~\citep[LOFAR:][]{LOFAR} and the Long Wavelength Array
~\citep[LWA:][]{LWA, LWA1}.  To better serve these needs we decided to
improve the survey sensitivity, reliability and uniformity by
re-reducing the data.

We have reprocessed all of the VLSS data from the archive to make new
maps and a new catalog, called the VLSS Redux (VLSSr).  New data
processing software including better RFI-removal algorithms, more
robust bright source peeling and smart-windowed cleaning were used to
improve the image quality. Bright isolated sources cataloged in the
VLSS served as a same-frequency starting sky model to improve the
ionospheric corrections applied to the data \citep{VLSSr1}.  Thousands
of sources observed multiple times in overlapping telescope pointings
were used to calculate a more accurate primary beam model.

The VLSSr corrects the radially-dependent flux errors present in the
VLSS, by using the newly calculated primary beam correction.  Compared
to the VLSS, the average map RMS noise level is reduced by $25$ per
cent to $\sigma \sim 0.1$ Jy beam$^{-1}$. The number of sources has
increased by $35$ per cent to 92,964.  The clean bias has been halved,
and is now $0.66 \times \sigma$, and the theoretical largest angular
size of $36$ arcmin is twice that of the VLSS.  Six previously
un-imaged fields are included in the VLSSr.

In this paper we describe the new VLSSr catalog and images.  The data
products are available on-line at
$<$URL:http://www.cv.nrao.edu/vlss/VLSSpostage.shtml$>$.

\section{The Data}

The VLSSr reprocessed all of the data from the original VLSS project.
The observations were made between 2001 and 2007, under VLA observing
programs AP397, AP441, AP452, and AP509. The sky was divided into a
roughly hexagonal grid of 523 pointing centers, at a spacing of $8.6$
degrees.  The bandwidth used was 1.56 MHz centered at 73.8 MHz.
Fields in the range $-10\degr < \delta < 80\degr$ were observed in the
VLA B configuration, while those at $ \delta < -10\degr$ and $\delta >
80\degr$ were observed in the BnA configuration to compensate for beam
elongation at low elevations. Further details of the observational
setup are given in \citet{VLSS1}.

Observations were made of each pointing using multiple 15-25 minute
scans, spread out over several hours to improve the hour angle
coverage.  Some fields were observed more than once, with days, months
and occasionally years separating the observations.  With a few
exceptions, fields were observed for a minimum of 75 minutes.  Longer
observations were made as needed in an effort to improve images for
pointings which did not meet the target quality criteria of $\sim0.1$
Jy beam$^{-1}$ RMS noise after preliminary reduction of the first 75
minutes of observations.

The observations made in 2006 and 2007 during the VLA upgrade slowly
decreased in quality due to decreasing numbers of antennas with
receivers, and most of the data were further corrupted due to changes
in the system itself.  These data were not included in the VLSS
catalogs and images.  We were unable to recover them and they had to
be left out of the VLSSr as well.

\section{VLSSr Data Reduction}
\begin{table*}
\begin{center}
\label{tab:fields}
\begin{tabular}{l c c c l}
\hline
Field & RMS noise (Jy beam$^{-1}$) & Dynamic Range & Reduction Method &
Notes \\
\hline
0000-041 & 0.0889 & 176 & 3 & \\
0000+041 & 0.0803 & 197 & 3 & \\
0000-123 & 0.0740 & 136 & 2 & \\
0000+123 & 0.0678 & 235 & 3 & \\
0000-208 & 0.0717 & 179 & 3 & \\
0000+208 & 0.0642 & 160 & 2 & \\
0000-298 & 0.0674 & 293 & 3 & \\
0000+298 & 0.0675 & 163 & 3 & \\
0000+398 & 0.0724 & 132 & brt, 3 & Cassiopeia A subtracted \\
0000+517 & 0.0815 & 4126 & brt, 3 & Cassiopeia A subtracted \\
0000+639 & 0.1628 & 4207 & brt, 2 & Cassiopeia A subtracted \\
0000+758 & 0.0662 & 265 & brt, 3 & Cassiopeia A subtracted \\
\hline
\noalign{\smallskip}
\end{tabular}
\caption{Notes on Individual Pointings.  For the Method column, ``2''
  indicates 2nd order Zernike fits, ``3'' indicates 3rd order Zernike
  fits, ``nvss'' indicates that the NVSS was used as a calibrator
  catalog, ``brt'' indicates bright source subtracted, with the source
  indicated in the notes column, ``peel'' indicates sources with peaks
  greater than 10 Jy beam$^{-1}$ were peeled, ``RFI'' indicates that
  more stringent RFI removal criteria were used.  The full table is
  available online at $<$URL:http://mnras.oxfordjournals.org/lookup/suppl/doi:10.1093/mnras/stu256/-/DC1$>$.}
\end{center}
\end{table*}

\begin{figure}
\noindent\includegraphics[width=20pc]{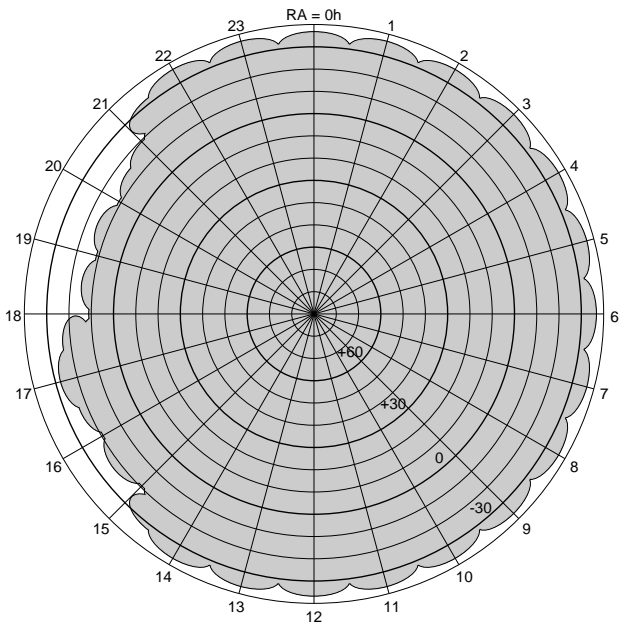}
\caption{A radial representation of the sky, with areas imaged in the
  VLSSr shaded. The circles represent declination in $10\degr$
  increments running from $\delta = 90\degr$ at the center to $\delta =
  -40\degr$ at the edge.  The radial lines represent hours of RA,
  increasing clockwise from 0h at the top to 23h. \label{fig:SkyArea}
}
\end{figure}

We describe here the data processing steps used to create the
VLSSr. More detailed descriptions of the algorithms used can be found
in ~\citet{VLSSr1}.

\subsection{Initial Calibration and Editing}
We performed initial bandpass and complex flux calibration using
observations of Cygnus A and a model of that source which had been
placed onto the \citet{Baars} flux scale. For each field, any data
points which were more than twice the total flux in that field were
clipped, typically removing $5--10$ per cent of the data. 

The data were then averaged in frequency to create 12 channels of
$\sim 120$ kHz width.  The frequency averaging was done to increase
the subsequent processing speed.  This corresponds to a reduction in
peak brightness and an increase in radial source width of about $5$
per cent at a radius of $\sim300$ synthesized beams, or $\sim 6\fdg3$
for our synthesized beam width of $75$ arcsec.  This is greater than
the VLA 74 MHz primary beam half-power radius of $\sim5\fdg6$, so peak
flux errors introduced by the frequency averaging should be less than
$5$ per cent.

These preliminary data reduction steps were performed in an older
version of the Astronomical Image Processing System ~\citep[31DEC03
  AIPS:][]{AIPS}, using the run-files and specialized reduction tasks
developed for the VLSS.  They are thus identical to the processing for
the original survey. However, we did not do the additional
RFI-excision tasks described in the VLSS paper \citep[sections 4.2.2
  and 4.2.4,]{VLSS1}.

\subsection{RFI-removal and Imaging for most fields}
The data were moved into the Obit data reduction package \citep{OBIT}
for further processing.

An initial imaging was performed (Obit task: IonImage), using
field-based calibration ~\citep{Cotton2004SPIE} to correct for the
errors introduced by the ionospheric phase-screen across the field of
view.  Isolated sources with $S > 2.5$ Jy beam$^{-1}$ measured in the
original VLSS survey were used as calibrators.  Solutions were
calculated at 1 minute intervals and times with poor solutions were
removed \citep{VLSSr1}.

Each pointing image covers a central region with a $7\fdg5$
radius.  The images are built up from hundreds of small facet images,
with sizes chosen by algorithms in Obit to avoid smearing caused by
3-D projection effects in the wide-field imaging
\citep{Cornwell1992,Cotton2004SPIE}.  The criteria used is that the
imaging phase error should be limited to 0.1 radian; for the VLSSr the
resulting facet diameters range from $\sim 0\fdg5 - 1\degr$, depending
on the maximum baseline and $w$ present in the data.  In addition to
the central region, sources out to a radius of $60\degr$ with
primary-beam corrected brightnesses greater than $2.5$ Jy beam$^{-1}$
were identified from the VLSS and imaged using small outlier facets.

The clean components found in the first imaging step were subtracted
from the data, leaving a residual data set.  This residual data set
was used to estimate and remove the RFI ~\citep[Obit tasks: RFIFilt,
  AutoFlag, ][]{VLSSr1, CottonRFIMemo}, with the resulting flags and
subtractions applied to the original (non-residual) data.  The
RFI-corrected data was then re-imaged with IonImage.

During this final imaging step, bright sources were temporarily
subtracted from the data set to minimize their sidelobe impact. The
data were imaged, and sources with a peak $S_{\nu} > 25$ Jy
beam$^{-1}$ were identified. For each of these, a residual data set
with all other field sources removed was generated from the initial
imaging and ionospheric calibration models.  The strong source was
centered on a pixel ~\citep{CottonUson2008}, and self-calibrated to
create an accurate model, which was then distorted by the ionospheric
calibration model and subtracted from the original data.  This
``peeling'' process was repeated for each bright source, and is
described in more detail in ~\citet{VLSSr1}.  The source-subtracted
data were then re-imaged and the saved bright source components were
restored to the final clean maps with the other clean components.

The B- and BnA- configuration data were processed identically, except
that the latter were given an upper UV-limit of $6 k\lambda$ to make
the angular resolution coverage more equal near the border between B
and BnA configuration coverage.  A round, $75$ arcsecond diameter
restoring beam was used in both cases.  Images were made with $15$
arcsecond pixels, which gives 5 pixels across the beam to provide
proper sampling and avoid discretization errors.

All of the data for each pointing center were processed at least
twice; once with a 2nd order and once with a 3rd order Zernike
polynomial to correct for the ionospheric phase-screen.  The better
map was kept, based on the criteria of higher dynamic range and total
flux in the field.  Visual inspection was then made of all final
images.  Images which had obvious calibration or other processing
errors were flagged for further processing, and/or were down-weighted
in the final image mosaics.

As described for the original survey \citep{VLSS1}, most maps
exhibited evidence of residual large-scale ionospheric shifts when we
compared the VLSSr and NVSS positions.  We calculated and applied
image-plane Zernike polynomial corrections for the shifts.  Because of
the large number of sources in the final images, we were able to fit
2nd, 3rd, and 4th order polynomials to the image shifts for all
images.  Images with the smallest average residual RMS to the fits
were kept.

\subsection{RFI-removal and Imaging for fields near bright sources}
The brightest, most extended, sources on the sky (such as Cygnus A)
can have flux densities that are hundreds or even thousands of
Janskys.  Because the primary antenna beam at 74 MHz for the VLA is so
large, these can dominate the flux in an observation, even at pointing
radii larger than the primary beam half-power point.  For example, the
primary beam pattern is attenuated by 15 dB at a pointing radius of
$\sim 11\fdg5$ \citep{VLA74}.  Cygnus A has a nominal flux
density on the ~\citet{Baars} scale of $S_{74} \sim 17$ kJy;
attenuated by 15 dB, this is still $\sim 550$ Jy.

The ionospheric calibration we perform relies on Zernike polynomials
which are unconstrained at radii beyond the beam half-power point.
Sources further out in the beam are poorly calibrated and may not be
completely removed during the standard initial imaging processing
step.  For a moderately weak source this effect is negligible, but for
bright sources such as Cygnus A, the remaining signal can dominate the
``residual'' signal for that pointing.  This makes the subsequent
RFI-removal step ineffective.

In order to avoid this issue, these sources can be removed before any
processing and then recombined later with the field images.
Observations centered within $20\degr$ of Cygnus A and Cassiopeia A,
and within $10\degr$ of Virgo A, Hydra A, Hercules A, and the Crab
were processed using this technique.

For each bright source, the data from the observations where it lay
closest to pointing center were self-calibrated, and a small image was
made at the position of the bright source.  These images were $\sim
6.4$ arcmin across for Virgo, and $\sim 3.2$ arcmin across for
the other 5 sources.

For each observation within the stated radius of the bright source, a
preliminary Zernike polynomial phase correction was calculated for
each time interval.  The bright source model was distorted by the
inverse of this phase correction at its position and subtracted from
the data.  The remaining data were then imaged, RFI corrected and
re-imaged as described previously.  The bright source-subtracted
images were compared to the images made by the regular reduction
method and the best final image was kept. The self-calibrated images
of the bright sources were saved and placed onto the final survey
images during the mosaic step described in Section 3.5.

\subsection{Additional variations}

A small number of fields which had higher RMS noise in the VLSSr than
in the original VLSS, or, on inspection, had clear sidelobes from
bright sources, were singled out for further processing.  These fields
were reprocessed separately using one or more of three variations on
input parameters:

\begin{enumerate}
\item
Using the NVSS as a sky model for the ionospheric calibration.  This
was particularly necessary when the VLSS itself had no image for a
pointing, or had a very poor image. (7 fields, $~\sim1.3$ per cent)
\item
Peeling all sources with peak flux $> 10$ Jy beam$^{-1}$. (8 fields,
$~\sim1.5$ per cent)
\item
Using a more stringent cutoff in the RFI removal step. (2 fields,
$~\sim0.4$ per cent)
\end{enumerate}

Fields were re-processed with both 2nd and 3rd order Zernike
polynomial corrections, and/or bright source-subtracted processing as
above, and the best map kept.  The total number of fields affected by
this alternate processing was $\sim3$ per cent.

Table ~\ref{tab:fields} gives a list of each pointing, the RMS
sensitivity over the inner half of the image, the final dynamic range
(peak to RMS) value, and the reduction method used on the final map.
While the dynamic range for most images is of order 100, fields with
bright sources reach dynamic ranges of several thousand.  The images
are thus unlikely to be dynamic range limited except near the very
brightest sources such as Cygnus A.

\subsection{Mosaic Images}
The field images were combined onto a set of overlapping $17$
deg$^{2}$ maps with 15 arcsec pixels.  Each image was corrected for
the primary beam, truncated at the beam half-power point, and weighted
by the inverse of its RMS in the combination step.  A small percentage
($\sim15$ per cent) of the fields were down-weighted by an additional
factor of 0.10, because a visual inspection revealed that the image
quality was poor in some way that might not be adequately reflected in
the RMS value (eg. RFI structure, distorted sources, image artifacts,
poor source distribution).  These images thus contribute almost
nothing to the final mosaics in any area where another image was
available to use, but are present where they are the only data
available; this minimizes their potentially negative effect on the
survey, while maximizing the coverage area.

Fields observed with the BnA configuration in 2006 and 2007 were
affected by an error which led to images with almost no sources
visible regardless of noise level. Despite significant effort we were
unable to correct the problem. These 25 fields ($\sim5$ per cent of the
total) were completely excluded from the final survey mosaics.  The
affected fields are marked in Table ~\ref{tab:fields}.  The final sky
coverage of the survey is shown in Figure ~\ref{fig:SkyArea}.

\subsection{Cataloging}
A catalog of sources was extracted from the mosaic images using the
Obit task FndSou to fit Gaussians to peaks in the maps.  Sources were
searched down to 3.5 times the global RMS value in each mosaic image;
the catalog was then filtered to keep only sources at greater than
$5\sigma$ significance, with the RMS noise measured locally in a
region with a 60 pixel ($15$ arcmin) radius.

The final catalog contains 92,964 entries. Of these, roughly $3$ per cent
have no match in the NVSS within $60$ arcseconds of the fitted
position.  However many unmatched sources are actually components of
larger sources which may be fit differently in the two catalogs.

To get a more accurate count of sources with no NVSS counterpart, we
filtered out $\sim 90,000$ isolated sources.  These are sources with
no 2nd VLSSr component within $120$ arcsec, which is the maximum
allowed source size used in the fitting algorithm.  Of these isolated
sources, $2.2$ per cent have no NVSS counterpart within that same $120$
arcsec.  For the purposes of survey statistics, we consider these
false detections.

A $5\sigma$ VLSSr source, with peak brightness $S_{74} = 0.5$ Jy
beam$^{-1}$ and a spectral index steeper than $\alpha^{1400}_{74} =
-1.8$ would fall below the NVSS peak brightness limit of $2.5$ mJy
beam$^{-1}$~\citep{NVSS}.  So some fraction of the sources without
counterparts may be real, steep-spectrum sources.  The rest are
generally either noise bumps near $5\sigma$, sidelobes of bright
sources, or edge pixels which were corrupted during the mosaic
procedure.  All of these sources remain in the final catalog.  If
there is any doubt about a specific catalog entry of interest, it can
be verified by inspecting the final images.

\section{Primary Beam}
\begin{table}
\label{tab:beam}
\begin{tabular}{l c c c}
\hline
Model & type & polynomial coefficients & FWHM (deg) \\
\hline
VLSS & J$_{inc}$ & & 9.7 \\
AIPS & 3rd order$^{a}$ & -0.897e-3 & 11.9 \\
 ~ & ~ & 2.71 e-7 & ~ \\
 ~ & ~ & -0.242e-10 & ~ \\
VLSSr & 3rd order$^{a}$ & -1.051e-3 & 11.2 \\
 ~ & ~ & 4.28e-8 & ~ \\
 ~ & ~ & -5.38e-11 & ~ \\
\hline
\end{tabular}
\\ 
{\em $^{a}$}The 3rd order polynomial has the form: $S = 1 + c1x\Theta
+ c2x\Theta^{2} + c3x\Theta^{3}$, where c1, c2, and c3 are the listed
coefficients, and $\Theta$ is the distance from the center of
the beam.
\caption{Beam model parameters at 74 MHz}
\end{table}

\begin{figure}
\noindent\includegraphics[width=20pc]{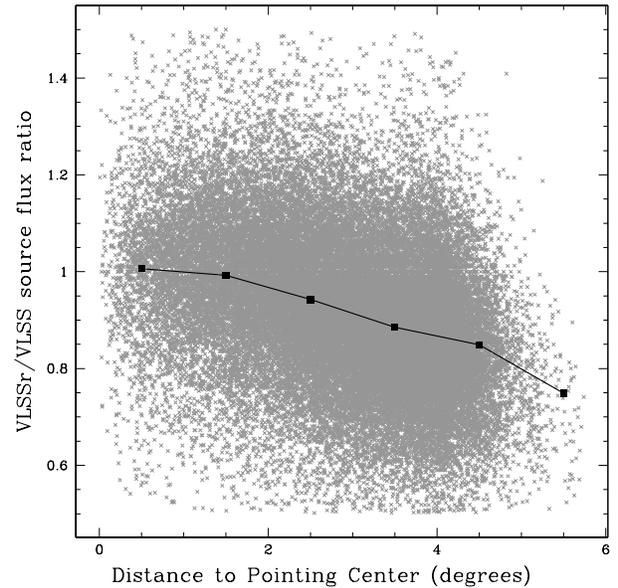}
\caption{ The ratios of the published VLSS flux densities corrected
  by a scaled $J_{inc}$ primary beam, to the VLSSr flux densities
  corrected with the the standard AIPS primary beam function, are
  plotted as a function of distance to the nearest pointing center.
  In order to avoid comparing resolved sources at two different
  resolutions ($80$ and $75$ arcsec for the VLSS and VLSSr
  respectively), only the $\sim 45000$ sources which are unresolved in
  both catalogs are included.  The mean value in 1 degree radial bins
  is overlaid on top.  The mean ratio for all sources is
  0.92. \label{fig:radius} }
\end{figure}

\begin{figure}
\noindent\includegraphics[width=15pc,angle=-90]{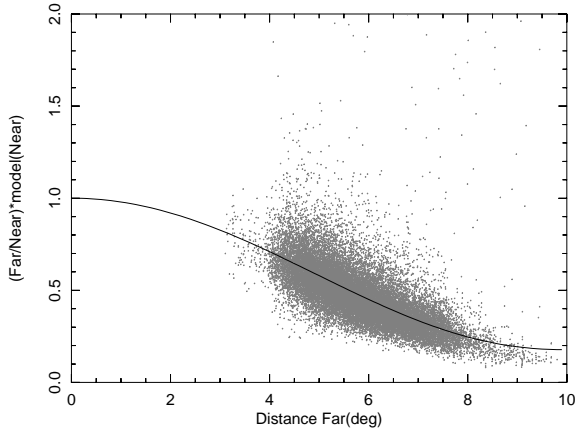}
\caption{The source pair data used to define a new beam polynomial are
  plotted as a function of radial distance to the pointing center of
  the source which is further from the center.  The polynomial fit is
  overlaid on the data. \label{fig:polynomial} }
\end{figure}

During verification of the final catalog, a flux discrepancy between
the previous VLSS and new VLSSr measurements was discovered.  We
compared unresolved sources to eliminate the effect of the different
clean beam sizes and found that the VLSSr sources had, on average,
roughly $10$ per cent lower measured flux than the sources in the
VLSS. Further investigation revealed that the flux deficit was not
present in the field maps, before primary beam corrections were made.
Re-mosaicking the original VLSS pointing maps using the Obit software
also failed recover the published fluxes.  We binned the sources by
radial distance from pointing center, and found there was a clear
increase in flux deficit as radius increased (see Figure
~\ref{fig:radius}).  This suggested a discrepancy between the primary
beam correction used in the previous reduction and the one we were
using for the VLSSr.

The standard beam correction for 74 MHz is a polynomial function of
$\Theta$, where $\Theta$ is defined as the angular offset from the
center of the primary beam.  It has a full-width at half maximum
(FWHM) of $11\fdg 8$ ~\citep{VLA74}.  However, looking at the
pipeline code used to reduce the original VLSS, that survey instead
used a specialized AIPS-task, WATE, which scaled a J$_{inc}$
function\footnote{defined as $J_{inc}(t) = {J_{1}(t)\over{t}}$, where
  $J_{1}(t)$ is a Bessel function of the first kind} from 1.47 GHz to
74 MHz to get a beam with a FWHM of only $9\fdg7$.

The pointings of the VLSSr overlap sufficiently to provide a source
sample for derivation of new primary beam parameters.  We used sources
observed at multiple pointing radii in the survey data itself to fit a
new polynomial for the primary beam.  We cataloged all of the pointing
images at local $5\sigma$ (using $\sigma$ measured within a $15$
arcmin radius of each source), and then searched for isolated sources
with no companions within $120$ arcsec, which appeared in multiple
fields; roughly 92,000 source pairs were identified.  To minimize
noise the sample was then filtered to include only pairs with both
source flux densities $> 1$ Jy and both pointing radii $< 6$ degrees
(roughly within the expected FWHM of the beam), and with flux density
ratios between 0.3 and 3.5 (to minimize the effect of sources with one
severely erroneous measurement).  The bright source Cygnus A was also
discarded, leaving approximately 10,500 source pairs.  We fit a third
order polynomial to the ratios of the two flux densities and the two
radii for each pair.  The fit was weighted by the sum of the flux
densities of each pair.

Figure ~\ref{fig:polynomial} plots the ratio of source fluxes times
the model flux for the source nearest the pointing center as a
function of the distance to the source farthest from the pointing
center.  The fitted polynomial is overlaid.  Table ~\ref{tab:beam}
summarizes the main parameters of the VLSSr beam, the standard AIPS
beam, and the VLSS beam.

\begin{figure*}
\includegraphics[width=40pc]{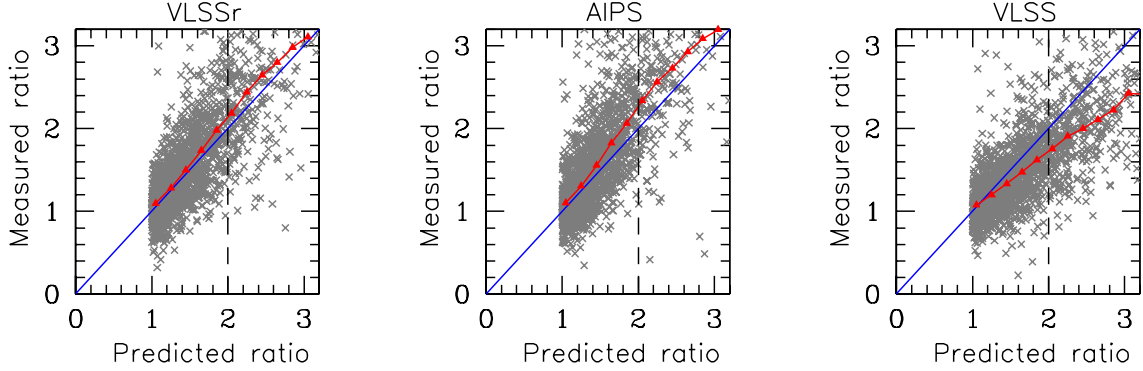}
\caption{Each panel shows a comparison of the predicted flux ratio at
  two different pointing radii vs. the measured flux ratio at the same
  radii for a given source.  The new VLSSr polynomial fit is on the
  left, the standard VLA 74 MHz polynomial from AIPS is in th middle
  and on the right we show the J$_{inc}$ function used for the VLSS.
  The blue diagonal line indicates a match between model prediction
  ratio and measured ratio, while the red triangles indicate the mean
  measured values.  A model ratio of two (indicated with the dashed
  line) is roughly equivalent to the half power point of the
  beam. \label{fig:beam} }
\end{figure*}

In each panel of Figure~\ref{fig:beam} we plot the ratio of the source
fluxes measured at two different pointing radii as a function of the
predicted source flux ratio at those two radii for one of the model
beams. The ratios were calculated so that they are always $>1$.

All three models show reasonable agreement with measured values at a
predicted flux ratio of 1; this is expected as the sources in each
pair at that ratio must lie at close to identical pointing-center
radii.  At larger predicted flux ratios, corresponding to greater
radial differences in the positions, the J$_{inc}$ function used by
the VLSS clearly over-predicts the measured flux density ratio.  The
use of the J$_{inc}$ function to correct the published VLSS maps thus
introduced a radially dependent flux density error.

The standard AIPS polynomial and the newly calculated VLSSr polynomial
both show good agreement between the predicted and measured flux
density ratios at predicted ratios between 1 and 1.5.  Somewhere
between predicted ratios of 1.5 and 2, both polynomials start to
under-predict the ratio as measured in the data.  This problem is more
pronounced in the standard AIPS polynomial, particularly at predicted
ratios greater than 2.  Thus these models will under-predict the flux
density of sources at large pointing radii.

A flux density ratio of 2 corresponds, to zeroth order, to a pair of
measurements with one at pointing center and one at the beam radius at
full-width half power.  Thus both the standard and VLSSr polynomials
should be used with caution outside the half-power point of the
beam. For this reason we have limited the VLSSr mosaics to data
within a radius equivalent to the beam half-power point of each
pointing center.

We are ignoring the known but un-characterized asymmetry of the 74 MHz
VLA primary beam \citep{VLA74}.  Because the pointing images combine
snapshots taken over a range of hour angles, we did not feel our
images were suitable to investigate the asymmetry.  However, it may
account for some of the scatter in the plots.

The standard reduction tasks within Obit have been adjusted to include
the new VLSSr fitted polynomial beam parameters for 74 MHz, and the
fitted beam was used on the final survey products.

\section{Survey Verification}
\subsection{Position Errors}
The global position offset and errors were measured by comparing
measured positions for VLSSr sources to their NVSS positions,
following the method previously described in \citet{VLSS1}.  We
defined a sample of 1284 sources that are weaker than our field
calibrators ($S < 2.5$ Jy) and stronger than $25\sigma$.  From these
we measure a global position offset of $\delta RA = -0.77$ arcsec
and $\delta Dec = 0.31$ arcsec.  Catalog entries have been
corrected for this bias.

The global RMS position error is $3.3$ arcsec in RA and $3.5$
arcsec in declination.  Positions of individual sources may also
be in error due to the Gaussian fitting technique we use, as described
in \citet{VLSS1} and \citet{CondonGauss}.  For each catalog source,
the calculated Gaussian errors are added in quadrature with the global
RMS position errors to derive the reported catalog position errors.
However, true position errors for any individual source in the VLSSr
catalogs will be dominated by residual errors from the ionospheric
corrections.  These errors are not suited to a global correction as
they will vary greatly from location to location, but may be on the
order of tens of arcseconds.

\subsection{Flux Density Errors}
In order to estimate the uncertainty in measured flux densities for
sources in the VLSS, we imaged all of the individual 20 minute
snapshots for each pointing, and cataloged the sources in them at a $>
5\sigma$ significance.  This gave us multiple peak flux density
measurements for each source.  We then compared the source peak flux
density to its mean for each source.  We find an average peak flux
error of $12\%$ for point sources, and $15\%$ for extended
sources.  We have included the point source error in the reported
catalog source error values.

The reported errors in the catalog also include the errors introduced
by the Gaussian fits, described in detail in \citep{VLSS1}.

\subsection{Flux Scale}\label{fluxscale}
\begin{table}
\label{tab:scale}
\begin{tabular}{l c c c c}
\hline
Source & Model order & Predicted & raw VLSSr & ratio \\
 & & Jy beam$^{-1}$ & Jy beam$^{-1}$ & \\
\hline
3C48 & 3rd & 77 & 69 & 1.12 \\
3C147 & 3rd & 52 & 52 & 1.0 \\
3C196 & 2nd & 133 & 121 & 1.10 \\
3C286 & 3rd & 31 & 28 & 1.11 \\
3C295 & 4th & 132 & 121 & 1.09 \\ 
3C380 & 1st & 133 & 112 & 1.19 \\
\hline
\end{tabular}
\caption{Scaife \& Heald Source model parameters}
\end{table}

\begin{figure}
\noindent\includegraphics[width=20pc]{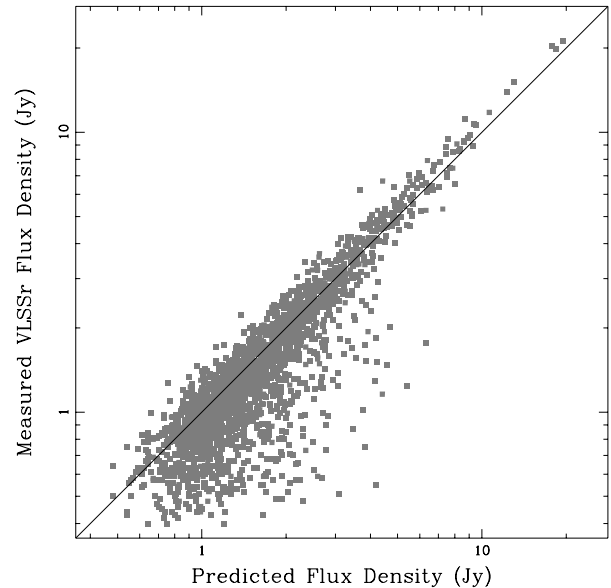}
\caption{A comparison of predicted and measure flux density for 2096
  isolated, unresolved VLSSr sources with matches in the 6C and 8C
  catalogs.  The predicted flux densities were calculated at 73.8 MHz
  by interpolating between the 6C (151 MHz) and 8C (38 MHz) flux
  densities. The line indicates a ratio between predicted and measured
  flux densities of unity. \label{fig:8C6Cflux} }
\end{figure}

During the course of investigating the primary beam-model, we also
tested our processing steps to see if they introduced any flux scale
bias.  We found that, with the exception of fields with extreme RFI,
the RFI filtering and flagging steps were reducing source fluxes in
the maps by $\sim5$ per cent.  Tests run independently on the filtering
vs. the flagging step showed that both steps contribute to the flux
loss.  

The original VLSS is assumed to lie on a Baars et al. scale
\citep{Baars}, because the primary calibration is tied to the Baars et
al. model for Cygnus A.  While some comparisons were made to existing
source fluxes (notably to K{\"u}hr et al. models \citep{Kuhr} and the
8C/6C catalogs \citep{8C,6C}), the conclusion made was that no flux
bias existed within the flux errors of the catalogs used
\citep{VLSS1}.  However the K{\"u}hr et al. models are not always reliable
at low frequencies (particularly for weaker sources) when there are
few existing low frequency measurements in the literature, and the
8C/6C catalogs are not themselves on the Baars et al. scale.

When we investigated the existing low frequency catalogs to generate
expected fluxes at 74 MHz to measure the flux bias in the VLSSr, we
found that the majority had been placed on the \citet[RCB]{rcb73}
scale.  This scale has the advantage over the Baars et al. scale that
it is based on low frequency flux measurements and thus consistent
down to the lowest frequencies.  The Baars et al. scale is tied to
models of Cas A and Cygnus A which are not accurate at very low
frequencies.

We compared the model predictions of all six bright 3C sources found
in \citet{ScaifeHeald2012} to the integrated flux densities in the
VLSSr to calculate an average flux density correction for our survey,
and place it onto the RCB scale.  A summary of the predicted and
measured integrated flux densities for the six sources is given in
Table ~\ref{tab:scale}.  The average correction factor was $1.10x$.
This scaling was applied to the VLSSr image mosaics before cataloging.
When using the VLSSr in conjunction with higher frequency, Baars et
al. scale catalogs, the scaling can be reversed, however, we suggest
that the assumed uncertainty in the flux measurement (discussed in
Section 5.2) should be increased by $\sim 5$ per cent to reflect the
known biases of the reduction method as discussed above.

In Figure ~\ref{fig:8C6Cflux}, we compare VLSSr flux densities to
those predicted by interpolating between the 6C ~\citep[151 MHz:
][]{6C} and 8C ~\citep[38 MHz: ][]{8C} catalogs. 

The 8C beam is $4.5\ \times 4.5 csc(\delta)$ arcmin; at a
declination of $\delta = 60\degr$ this corresponds to $4.5 \times 5.2$
arcmin.  The 6C beam is only slightly smaller, at $4\ \times 4
csc(\delta)$ arcmin.  However, the VLSSr beam is only $1.25$
arcmin.  In order to compare flux density between the three
catalogs, we constructed a sample of isolated, unresolved sources.  We
require that all sources be present in all three catalogs.

For the 8C catalog we required that sources have the descriptor ``P''
for point source.  We used the integrated flux density when available,
and peak intensity when it was not (peak intensity for an unresolved
point source should be the same as integrated flux density).  For the
6C sources, we required that they not be marked as part of a source
complex. As with the 8C catalog, we used integrated flux density when
available and peak flux density for the remaining sources.  

For the VLSSr we used the deconvolved total source flux density, which
includes the clean bias correction.  We required that the VLSSr have
only one source entry within a $6$ arcmin radius of the reported 8C
source position to ensure that we were comparing single isolated
sources.  Finally we required that the source be unresolved in the
VLSSr, with only upper limits for the deconvolved source major and
minor axes.

\begin{figure}
\noindent\includegraphics[width=20pc]{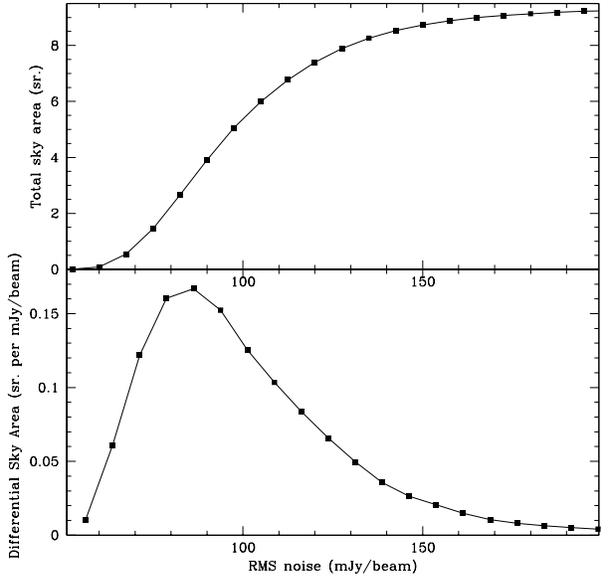}
\caption{{\it top} Total sky area in steradians (y-axis) at or below a
  given RMS noise level (x-axis).  The mean noise level is 0.1 Jy
  beam$^{-1}$. Values have been measured from residual
  (source-subtracted) images. {\it bottom} The differential noise level
  in steradians per Jy beam$^{-1}$.  \label{fig:SkyNoise} }
\end{figure}

For the sample of 2096 sources thus selected the median ratio of
measured VLSSr to predicted flux density is $0.92$.  There is a trend
for the VLSSr to be below the prediction at lower flux densities and
above the prediction at higher flux densities.  Thus for the 231
sources with $S_{74} \geq 3$ Jy, the median ratio of measured to
predicted flux densities is $1.07$, while for the 674 sources with
$S_{74}< 1$ Jy, the median ratio of measured to predicted flux density
is only $0.72$.  It is uncertain what might cause this flux-dependent
scale variation, however it is clear that the weakest sources, in
particular, are not reliable in this comparison.  If we limit the
sample to sources with $S_{74} > 1$ Jy, the median ratio of measured
to predicted flux density is $0.97$.

\section{Sky Area Imaged}

In Figure ~\ref{fig:SkyArea}, we show the sky coverage of the
VLSSr. The survey covers roughly $3\pi$ sr, equivalent to the intended
area of the original observations.  The resulting images completely
cover declinations $\delta > -10\degr$ for $18^{\rmn{h}} < RA <
21^{\rmn{h}}$ and $\delta > -20\degr$ for $15^{\rmn{h}} < RA <
18^{\rmn{h}}$.  The lower declination fields at these right ascensions
were corrupted and the data could not be imaged.  The remainder of the
sky is covered for all declinations $\delta > -30\degr$, with a
scalloped edge extending to $\delta \simeq -36\degr$.

\section{Sky Noise Properties}

The average RMS noise over the entire survey area is $\sigma = 0.1$ Jy
beam$^{-1}$. Figure ~\ref{fig:SkyNoise} indicates how much sky was
observed at a given RMS noise level.  The majority of the survey has
an RMS within $\pm 0.03$ Jy beam$^{-1}$ around the median noise level
of $0.095$ Jy beam$^{-1}$, with a long tail of larger RMS values.  This
tail is comprised largely of regions near very strong sources where
sidelobes raise the local noise level.  It also includes regions near
the edge of the survey area where there is no neighboring field to
improve the sensitivity, and regions located along the higher
temperature Galactic plane.  Finally it includes some areas where the
ionospheric calibration was poor due either to a lack of calibrator
sources or unusually bad ionospheric weather.

\begin{figure}
\noindent\includegraphics[width=20pc]{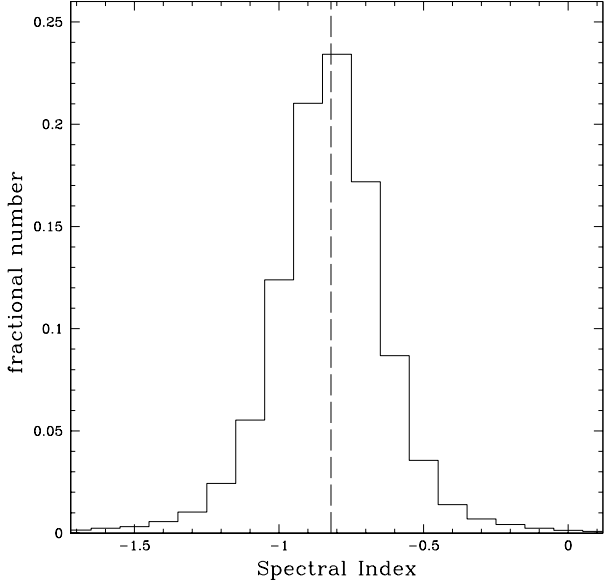}
\caption{Histogram showing the spectral index distribution for sources
  from the VLSSr compared to the NVSS (1400 MHz, black).  The median
  spectral index, $\alpha^{1400}_{74} = -0.82$ is shown with a
  dashed line. \label{fig:sicompare} }
\end{figure}
\section{Source Spectral Indices}

We calculated an average spectral index to 1400 MHz, using the NVSS
~\citep{NVSS} for flux values.  We started with $\sim 90,000$ isolated
VLSSr sources with no additional source in a $120$ arcsecond radius,
and searched for NVSS matches within the same $120$ arcsecond radius.
After removing sources with more than one NVSS match, we were left
with 67,844 unique source pairs. We divided the VLSSr flux densities
by 1.1 to re-place them onto the default Baars et al. scale used by
the NVSS, and calculated spectral index values for each source.  The
median spectral index is $\alpha^{1400}_{74} = -0.82$.  The spread in
the derived median, represented by the semi-inter-quartile range is
$SIQR = 0.11$.

Figure ~\ref{fig:sicompare} shows the spectral index distribution,
which is reasonably symmetric.  The median agrees within the errors
with the median spectral index of $\alpha^{1400}_{325} \sim -0.9$
found between bright WENSS sources and the NVSS by \citet{Zhang2003},
and the value of $\alpha^{1400}_{74} \sim -0.72$ found for the bright
sample in a deep low frequency study of the XMM-LSS field
\citep{Tasse2006}.

\section{log N - log S}
\begin{figure}
\noindent\includegraphics[width=20pc]{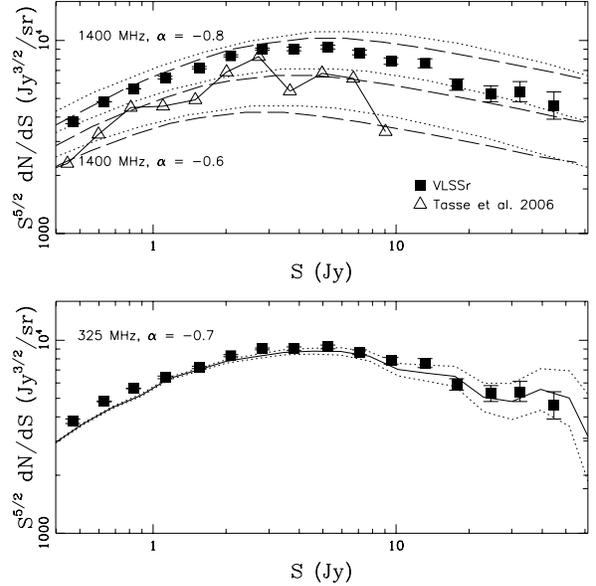}
\caption{Euclidean-normalized source counts for the VLSSr are shown,
  scaled onto the Baars et al. flux scale.  {\it Top} VLSSr counts
  (squares) are overlaid on curves from the 1.4 GHz source counts
  given in Condon (dotted line, 1984) and Seymour et al. (dashed line,
  2004).  The 1.4 GHz curves have been scaled to 74 MHz by spectral
  indices of $\alpha = -0.6, -0.7$ and $-0.8$.  We also plot values
  for 74 MHz counts taken from \citet[triangles, ][]{Tasse2006}. {\it
    Bottom} VLSSr counts are overlaid on a curve derived from WENSS
  sources \citep{WENSS} and scaled to 74 MHz by a spectral index of
  $\alpha = -0.7 \pm 0.3$. Errors in the WENSS counts are indicated by
  the dotted lines.  \label{fig:logNlogS} }
\end{figure}

\begin{table*}
\begin{minipage}{126mm} 
 \label{tab:logNlogS}
 \begin{tabular}{l c r c c c c}\hline 
Flux range & $\log_{10} S_c$ & Raw counts & Mean Area & Mean weight & Normalized counts \\ 
(Jy) & (Jy) &  & (sr) &  & ($10^3$ Jy$^{3/2}$ sr$^{-1}$) \\ 
\hline \hline 
0.39--0.53 & $-0.32$ & 5813 & $2.4$ & $0.17$ & $\phantom{1}3.7^{+0.1}_{-0.1}$\\
0.53--0.72 & $-0.20$ & 16490 & $5.6$ & $0.53$ & $\phantom{1}5.4^{+0.0}_{-0.0}$\\
0.72--0.98 & $-0.08$ & 18825 & $8.1$ & $0.85$ & $\phantom{1}6.1^{+0.0}_{-0.0}$\\
0.98--1.33 & $\phantom{-}0.05$ & 16129 & $9.0$ & $0.97$ & $\phantom{1}7.3^{+0.1}_{-0.1}$\\
1.33--1.81 & $\phantom{-}0.19$ & 11423 & $9.3$ & $0.99$ & $\phantom{1}8.0^{+0.1}_{-0.1}$\\
1.81--2.47 & $\phantom{-}0.32$ & 8332 & $9.3$ & $1.00$ & $\phantom{1}9.2^{+0.1}_{-0.1}$\\
2.47--3.36 & $\phantom{-}0.45$ & 5768 & $9.3$ & $1.00$ & $10.1^{+0.1}_{-0.1}$\\
3.36--4.57 & $\phantom{-}0.58$ & 3755 & $9.3$ & $1.00$ & $10.4^{+0.2}_{-0.2}$\\
4.57--6.22 & $\phantom{-}0.72$ & 2411 & $9.3$ & $1.00$ & $10.6^{+0.2}_{-0.2}$\\
6.22--8.46 & $\phantom{-}0.85$ & 1457 & $9.3$ & $1.00$ & $10.2^{+0.3}_{-0.3}$\\
8.46--11.5 & $\phantom{-}0.98$ & 864 & $9.3$ & $1.00$ & $\phantom{1}9.6^{+0.3}_{-0.3}$\\
11.5--15.7 & $\phantom{-}1.12$ & 504 & $9.3$ & $1.00$ & $\phantom{1}8.9^{+0.4}_{-0.4}$\\
15.7--21.3 & $\phantom{-}1.26$ & 260 & $9.3$ & $1.00$ & $\phantom{1}7.2^{+0.5}_{-0.4}$\\
21.3--29.0 & $\phantom{-}1.39$ & 138 & $9.3$ & $1.00$ & $\phantom{1}6.1^{+0.6}_{-0.5}$\\
29.0--39.5 & $\phantom{-}1.52$ & 95 & $9.3$ & $1.00$ & $\phantom{1}6.7^{+0.8}_{-0.7}$\\
39.5--53.7 & $\phantom{-}1.66$ & 45 & $9.3$ & $1.00$ & $\phantom{1}5.0^{+0.9}_{-0.7}$\\
53.7--73.0 & $\phantom{-}1.77$ & 24 & $9.3$ & $1.00$ & $\phantom{1}4.2^{+1.1}_{-0.9}$\\
73.0--99.4 & $\phantom{-}1.95$ & 7 & $9.3$ & $1.00$ & $\phantom{1}2.0^{+1.1}_{-0.7}$\\
99.4--135 & $\phantom{-}2.09$ & 12 & $9.3$ & $1.00$ & $\phantom{1}5.3^{+2.1}_{-1.5}$\\
135--184 & $\phantom{-}2.20$ & 6 & $9.3$ & $1.00$ & $\phantom{1}4.2^{+2.6}_{-1.7}$\\
184--250 & $\phantom{-}2.28$ & 2 & $9.3$ & $1.00$ & $\phantom{1}2.2^{+3.1}_{-1.4}$\\
250--341 & $\phantom{-}2.52$ & 1 & $9.3$ & $1.00$ & $\phantom{1}1.8^{+4.3}_{-1.5}$\\
341--464 & $\phantom{-}2.63$ & 3 & $9.3$ & $1.00$ & $\phantom{1}8.5^{+8.5}_{-4.6}$\\
464--631 & $\phantom{-}2.71$ & 2 & $9.3$ & $1.00$ & $\phantom{1}9.0^{+12.3}_{-5.8}$\\
631--858 & $\phantom{-}2.80$ & 1 & $9.3$ & $1.00$ & $\phantom{1}7.1^{+17.2}_{-5.9}$\\
858--1168 & $\phantom{-}2.97$ & 1 & $9.3$ & $1.00$ & $11.3^{+27.3}_{-9.4}$\\
 \end{tabular}
\caption{Source count data for the VLSSr.  Tabulated values are on the RCB flux scale.}
\end{minipage}
\end{table*}

\begin{table*} 
\begin{minipage}{126mm} 
\label{tab:logNlogSBaars}
\begin{tabular}{l c r c c c c}\hline 
Flux range & $\log_{10} S_c$ & Raw counts & Mean Area & Mean weight & Normalized counts \\ 
(Jy) & (Jy) &  & (sr) &  & ($10^3$ Jy$^{3/2}$ sr$^{-1}$) \\ 
\hline \hline 
0.39--0.53 & $-0.33$ & 9402 & $3.4$ & $0.27$ & $\phantom{1}3.8^{+0.1}_{-0.1}$\\
0.53--0.72 & $-0.20$ & 18049 & $6.6$ & $0.65$ & $\phantom{1}4.8^{+0.0}_{-0.0}$\\
0.72--0.98 & $-0.08$ & 18508 & $8.5$ & $0.90$ & $\phantom{1}5.6^{+0.0}_{-0.0}$\\
0.98--1.33 & $\phantom{-}0.05$ & 14324 & $9.1$ & $0.98$ & $\phantom{1}6.4^{+0.1}_{-0.1}$\\
1.33--1.81 & $\phantom{-}0.19$ & 10420 & $9.3$ & $1.00$ & $\phantom{1}7.2^{+0.1}_{-0.1}$\\
1.81--2.47 & $\phantom{-}0.32$ & 7535 & $9.3$ & $1.00$ & $\phantom{1}8.3^{+0.1}_{-0.1}$\\
2.47--3.36 & $\phantom{-}0.45$ & 5139 & $9.3$ & $1.00$ & $\phantom{1}9.0^{+0.1}_{-0.1}$\\
3.36--4.57 & $\phantom{-}0.58$ & 3244 & $9.3$ & $1.00$ & $\phantom{1}9.0^{+0.2}_{-0.2}$\\
4.57--6.22 & $\phantom{-}0.72$ & 2102 & $9.3$ & $1.00$ & $\phantom{1}9.2^{+0.2}_{-0.2}$\\
6.22--8.46 & $\phantom{-}0.85$ & 1232 & $9.3$ & $1.00$ & $\phantom{1}8.6^{+0.3}_{-0.2}$\\
8.46--11.5 & $\phantom{-}0.98$ & 706 & $9.3$ & $1.00$ & $\phantom{1}7.8^{+0.3}_{-0.3}$\\
11.5--15.7 & $\phantom{-}1.12$ & 430 & $9.3$ & $1.00$ & $\phantom{1}7.6^{+0.4}_{-0.4}$\\
15.7--21.3 & $\phantom{-}1.25$ & 212 & $9.3$ & $1.00$ & $\phantom{1}5.9^{+0.4}_{-0.4}$\\
21.3--29.0 & $\phantom{-}1.39$ & 120 & $9.3$ & $1.00$ & $\phantom{1}5.3^{+0.5}_{-0.5}$\\
29.0--39.5 & $\phantom{-}1.51$ & 77 & $9.3$ & $1.00$ & $\phantom{1}5.4^{+0.7}_{-0.6}$\\
39.5--53.7 & $\phantom{-}1.65$ & 41 & $9.3$ & $1.00$ & $\phantom{1}4.6^{+0.8}_{-0.7}$\\
\hline
53.7--73.0 & $\phantom{-}1.80$ & 13 & $9.3$ & $1.00$ & $\phantom{1}2.3^{+0.8}_{-0.6}$\\
73.0--99.4 & $\phantom{-}1.93$ & 8 & $9.3$ & $1.00$ & $\phantom{1}2.2^{+1.1}_{-0.8}$\\
99.4--135 & $\phantom{-}2.06$ & 11 & $9.3$ & $1.00$ & $\phantom{1}4.9^{+2.0}_{-1.5}$\\
135--184 & $\phantom{-}2.21$ & 6 & $9.3$ & $1.00$ & $\phantom{1}4.2^{+2.6}_{-1.7}$\\
250--341 & $\phantom{-}2.48$ & 1 & $9.3$ & $1.00$ & $\phantom{1}1.8^{+4.3}_{-1.5}$\\
341--464 & $\phantom{-}2.60$ & 4 & $9.3$ & $1.00$ & $11.3^{+9.2}_{-5.4}$\\
464--631 & $\phantom{-}2.73$ & 2 & $9.3$ & $1.00$ & $\phantom{1}9.0^{+12.3}_{-5.8}$\\
631--858 & $\phantom{-}2.93$ & 1 & $9.3$ & $1.00$ & $\phantom{1}7.1^{+17.2}_{-5.9}$\\
 \end{tabular}
\caption{Source count data for the VLSSr.  Tabulated values are on the Baars et al. flux scale.  Data above the line have been used in Figure ~\ref{fig:logNlogS}, while those below the line were excluded from the figure due to small source number counts. }
\end{minipage}
\end{table*}

We used the cumulative distribution of RMS noise seen in Figure
~\ref{fig:SkyNoise} to derive the differential source counts for the
92,964 sources in our catalog. For each source, we compute $A_{\rm
  max}$, the total area for which $F_p > 5\sigma$ (where $F_p$ is the
peak flux and $\sigma$ is the local RMS noise as computed in the
cataloging step, see Sec.~3.6).

The differential source count is given by

\begin{equation}
dN = {dF_i}^{-1} \sum^N_{j=1} \left(\frac{1}{A_{\rm max}}\right)_j
\end{equation}

where $N$ is the number of sources in a bin of width $dF_i$ (with
$F_i$ the deconvolved integrated flux as measured after applying the
clean bias to the images). For large $N$, the uncertainty on the
source count is given by

\begin{equation}
\sigma_{dN} = {dF_i}^{-1}  \left[ \sum^N_{j=1}\left(\frac{1}{A_{\rm max} }\right)_j^{2} \right]^{1/2}
\end{equation}
\citep{Windhorst85,Condon02}.

When the mean weight, $N^{-1} \sum A_{\rm max}$, is unity, we compute
the uncertainty using Poisson statistics \citep{Gehrels86}. The
results are tabulated in Table \ref{tab:logNlogS}. 

In order to compare our source counts to those found in other sky
surveys, we have scaled the VLSSr to the Baars et al. flux scale and
recalculated the counts (see Table \ref{tab:logNlogSBaars}.  In Figure
\ref{fig:logNlogS} we plot the results.  We have excluded VLSSr flux
bins with fewer than 20 sources from the figure.

In the top panel we have plotted our source counts and overlaid curves
taken from the 1.4 GHz source counts in \citet{Condon1984} and
\citet{Seymour2004}, which have been extrapolated to 74 MHz using a
range of assumed spectral indices.  At low and moderate flux values
our counts are in reasonable agreement with the extrapolated curve
using $\alpha^{1400}_{74} = -0.8$; this in turn agrees well with the
median spectral index found previously between the VLSSr and NVSS.  At
higher flux values we find a steeper drop-off than seen in the 1.4 GHz
curve.

\citet{Tasse2006} previously used a deep study of the XMM-LSS field to
calculate source counts at 74 MHz.  Their counts, based on $\sim 650$
sources observed at higher resolution than the VLSSr ($\Theta \sim 30$
arcsec) are overlaid on the top panel.  We have left off the errorbars
to simplify the display.  The Tasse curve falls below the VLSSr curve,
but, when compared to the 1400 MHz counts, it is consistent with the
lower mean spectral index ($\alpha^{1400}_{74} = -0.72$) they also
derive.

In the bottom panel the overlaid curve is derived from an analysis of
the 325 MHz WENSS catalog \citep{WENSS}, with the $1\sigma$ Poisson
uncertainty indicated by the dotted curves. We used the calculated
spectral index and scatter between VLSSr (on the Baars et al. scale)
and WENSS ($\alpha = -0.7 \pm 0.3$) to assign a 74 MHz flux to each
WENSS source. The source count that is obtained using this Monte Carlo
method is in good agreement with the VLSSr curve.

\section{VLSS vs. VLSSr}
\begin{figure*}
\noindent\includegraphics[width=20pc]{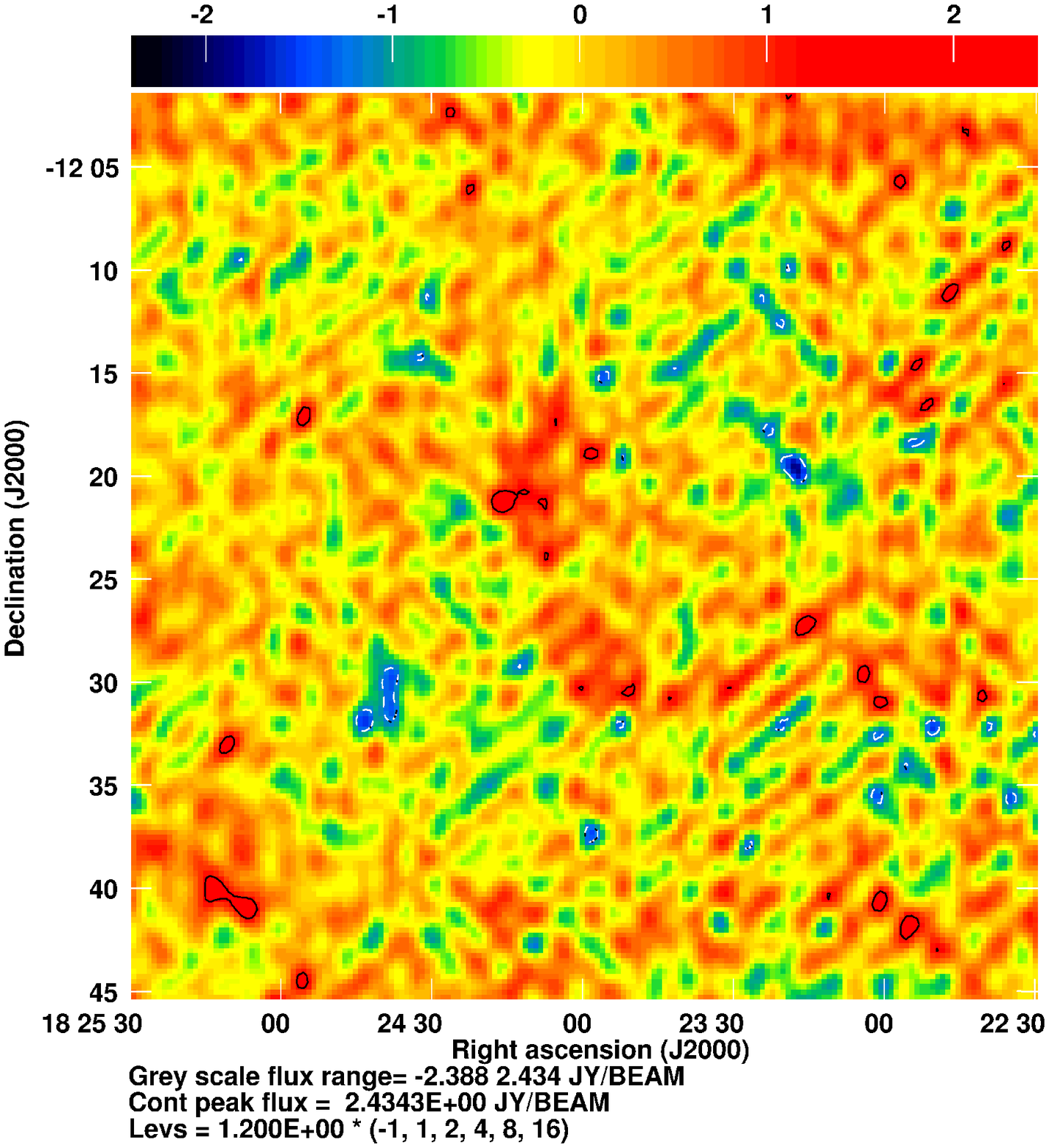}\includegraphics[width=20pc]{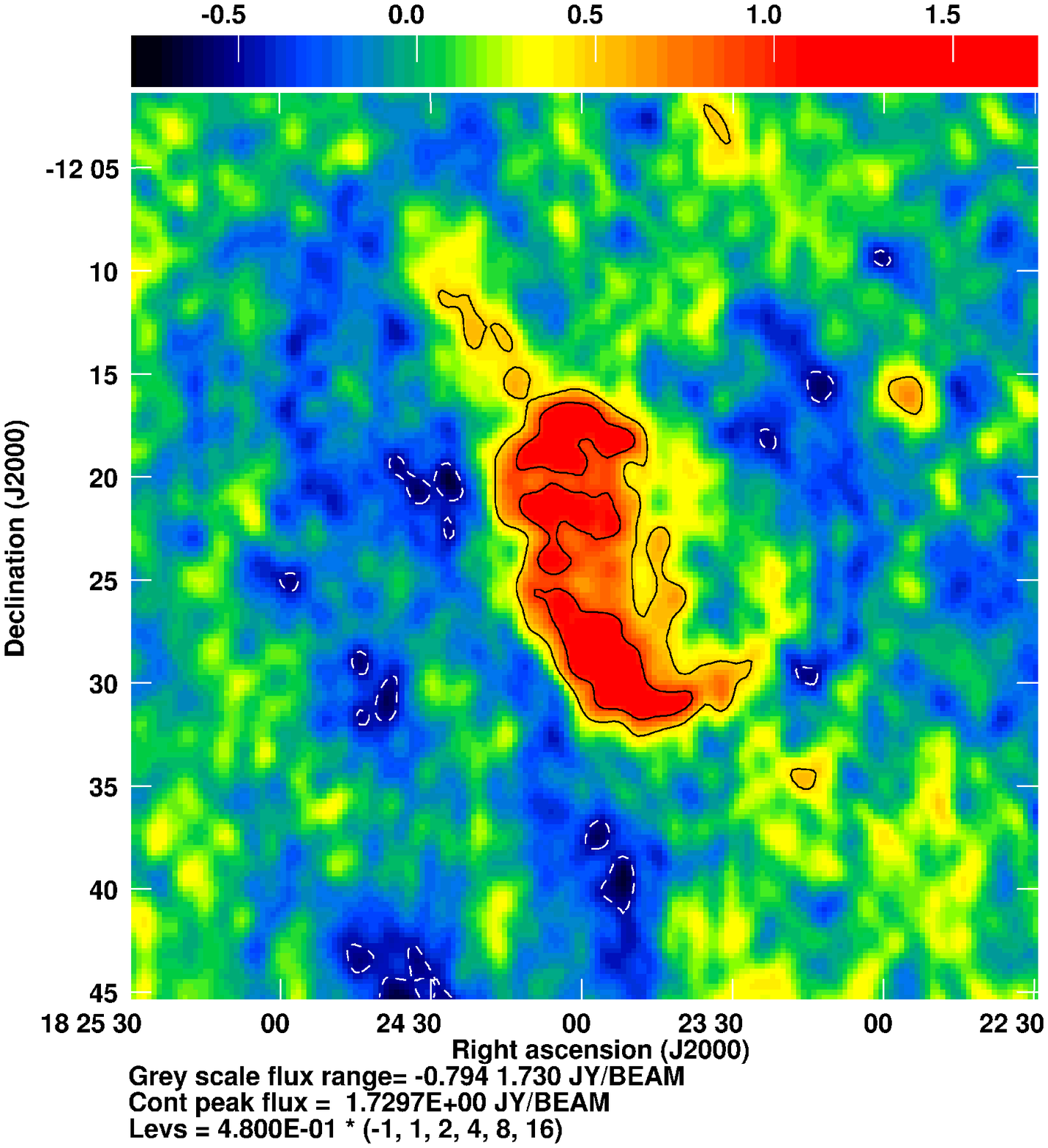}
\caption{A comparison of the supernova remnant Kes 63 in the VLSS (left) and VLSSr (right) processing. \label{fig:pretty} }
\end{figure*}

The VLSS catalog and images were limited by the available software and
computer processing power, as well as a lack of a same-frequency
initial sky model for the ionospheric calibration step.  The VLSSr
reduction took advantage of experience gained from the VLSS itself to
improve the final data products in many areas.  Figure
~\ref{fig:pretty} shows a sample extended source imaged by both
surveys, and Table ~\ref{tab:compare} summarizes the main survey
parameters for each.

The VLSSr uses a slightly smaller clean beam ($75$ arcsec) and pixel
scale ($15$ arcsec) compared to the VLSS.  We felt the smaller beam
was a more accurate reflection of the natural data properties, and the
pixel scale was lowered to provide 5 pixels resolution across the
beam.

The VLSSr processing took advantage of a new smart-windowing algorithm
to clean only areas in the images that have real sources and avoid
cleaning noise bumps \citep{VLSSr1, autoWindow}.  When noise is
cleaned, flux is subtracted from real sources, introducing a flux bias
which scales with the local RMS noise.  The point source clean-bias in
the VLSSr is $0.66\sigma$, where $\sigma$ is the locally measured RMS
noise in the image.  This is slightly less than half the clean-bias
from the VLSS.  The online catalog corrects for the clean-bias in the
deconvolved flux values.
Improved RFI-removal algorithms for the VLSSr allow the inclusion of
all short baselines present in the data.  In order to mitigate RFI
which tends to affect short baselines more than longer baselines, the
VLSS processing removed all baselines shorter than $200 \lambda$.  By
keeping the short baselines, the VLSSr doubles the theoretical largest
angular size (LAS) imaged to $36$ arcmin.  Short observations with
poor uv-coverage will reduce the actual angular size measured in the
maps.  The observations were spread out in hour angle as much as
possible to mitigate this effect; however it is probable that the
maximum LAS was not reached in all areas on the sky.

Because of the ionospheric phase calibration improvements from using a
same-frequency sky catalog, we were able to process all fields for the
VLSSr without doing any initial self-calibration.  In the VLSS this
step was necessary for many fields to be imaged at all.  While a small
number of VLSSr fields have visible calibration issues (distorted
source shapes or doubled sources are the most common) which may have
been ameliorated by the self-calibration step, we preferred to avoid
the hybrid phase treatment of self-calibration followed by
field-calibration.

The improvements in ionospheric calibration and RFI removal lead to
substantially better final images.  If we scale the VLSSr to the Baars
et al. flux scale to make a comparison, the average RMS noise has
decreased by a factor of $25$ per cent in the VLSSr to $\sigma = 0.09$
Jy beam$^{-1}$.  Higher-declination holes in the VLSS coverage area
have been filled in by the VLSSr, and sensitivity to large sources
(eg. Galactic supernova remnants), is greatly improved.

In the original VLSS survey, a comparison of 200 sources with $S_{74}
> 4$ Jy had an average ratio of measured VLSS flux to flux predicted
by extrapolating between the 6C and 8C catalogs of $\sim0.99$ with a
$10\%$ rms scatter \citep{VLSS1}.  By comparison, the same ratio is
$\sim 1.07$ with a $13\%$ rms scatter for the roughly 230 unresolved
sources in the VLSSr $S_{74}\geq 3$ Jy.  While it is clear that the
VLSS appears to be in better agreement with the prediction for these
bright sources, we note that no effort was made to ensure that the
VLSS sample was restricted to unresolved sources, so the comparison
included extended sources measured with very different beams.  The
VLSS result was also affected by the known position dependent flux
errors in that catalog.

The VLSSr rate of unmatched sources is higher than for the original
VLSS, with $2.2$ per cent of isolated VLSSr sources having no NVSS
counterpart within $120$ arcsec, compared to $\leq 1$ per cent in the
VLSS.  One reason is that the VLSS filtered the catalog to require a
higher significance for inclusion near bright sources in an effort to
reduce the inclusion of sidelobes from sources \citep{VLSS1}. Because
nearly half of the sources removed by this filter were real (had NVSS
counterparts), we chose not to use this filter for the VLSSr.

\begin{table}
\label{tab:compare}
\begin{tabular}{l c c}
\hline
Property & VLSS & VLSSr \\
\hline
Clean beam (FWHM) &  $80\arcsec$ & $75\arcsec$  \\
Pixel scale & $25\arcsec$ & $15\arcsec$  \\
Mean RMS (Jy beam$^{-1}$)$^{a}$ & 0.12  & 0.09 \\
Clean bias & $1.39\sigma$ & $0.66\sigma$  \\
Largest angular size & $18\arcmin$ & $36\arcmin$  \\
Total sky area (sr) & 9.43   & 9.38  \\
Sources detected ($\geq 5\sigma$) & $68,311$ & $92,964$  \\
Catalog non-match rate & $1\%$ & $2.2\%$ \\
\hline
\end{tabular}
\\ 
{\em $^{a}$}RMS is quoted using the Baars et al. scale for both.
On the \citet{ScaifeHeald2012} scale, the mean RMS of the VLSSr is
$\sigma \sim 0.1$ Jy beam$^{-1}$.
\caption{Comparison of VLSS and VLSSr Processing}
\end{table}

\begin{figure}
\noindent\includegraphics[width=15pc,angle=-90]{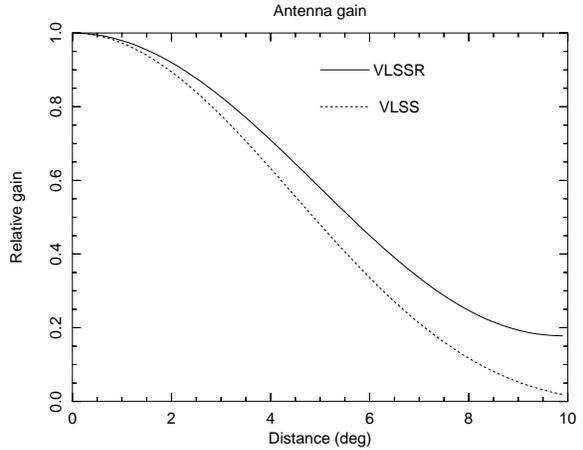}
\caption{The scaled J$_{inc}$ shaped-beam used to correct the VLSS
  vs. the fitted polynomial beam used for the VLSSr as a function of
  radius.  The two beams show a clear discrepancy which increases with
  increasing radius.  \label{fig:2beam} }
\end{figure}

A final major improvement is the use of a more accurate primary beam
correction for the VLSSr.  A comparison of the VLSS and VLSSr primary
beams is shown in Figure ~\ref{fig:2beam}.  The primary beam
correction for the VLSS over-corrected source fluxes, with the size of
the error changed with radial distance to closest pointing center.  As
a result, while the global flux scale for the VLSS looked quite
reasonable when compared to literature spectra and source flux values,
measurements of a given source could be wrong by a factor which varied
by source location.  Correction of the primary beam shape for the
VLSSr provides substantially more reliable and uniform source flux
measurements and removes the radially dependent flux errors.

\section {Conclusions}

The VLSSr provides a more accurate and more detailed map of the sky at
74 MHz than the original VLSS.  The improved sensitivity is reflected
in greater detail on images of faint structures.  The increased
angular scale allows measurements of supernova remnants and other
large sources not visible in the final VLSS images.

We have derived an improved primary beam model for the 74 MHz VLA,
based on sources observed multiple times in adjacent pointings.  The
new model is implemented in the Obit software reduction package.  We
have used it to correct a substantial primary beam error in the
original VLSS.

In keeping with other low-frequency catalogs we have placed the survey
onto the RCB scale \citep{rcb73}, using the source models from
\citet{ScaifeHeald2012}.  Those who wish to compare VLSSr data to
higher frequencies can re-place the survey products onto the
\citet{Baars} scale, by dividing all fluxes by a factor of 1.1.

All data (images and catalog) are publicly available at
$<$URL:http://www.cv.nrao.edu/vlss/VLSSpostage.shtml$>$.

\section*{Acknowledgments}
Basic research at the Naval Research Lab is supported by 6.1 base
funds.  We thank E. Polisensky for help with the graphics.  Part of
this research was carried out at the Jet Propulsion Laboratory,
California Institute of Technology, under a contract with the National
Aeronautics and Space Administration. The National Radio Astronomy
Observatory is a facility of the National Science Foundation operated
under cooperative agreement by Associated Universities, Inc.

\bibliography{vlssr.astroph}
\bibliographystyle{mn2e}

\clearpage

\label{lastpage}

\end{document}